\documentclass{osa-article}
\usepackage{siunitx}
\usepackage{amsmath}
\usepackage[version=4]{mhchem}

\journal{osajournal}

\articletype{Research Article}

\begin{document}

\title{The role of beam homogeneity in mechanical coupling evaluation of laser ablation-generated impulse}

\author{Jacopo Terragni,\authormark{1,*} Pietro Battocchio,\authormark{1}, Nicola Bazzanella,\authormark{1} Michele Orlandi,\authormark{1} William J. Burger,\authormark{2} Roberto Battiston,\authormark{1,2} and Antonio Miotello\authormark{1}}

\address{\authormark{1}University of Trento, Department of Physics, Via Sommarive 14, 38123, Povo (TN), Italy\\
\authormark{2}Trento Institute for Fundamental Physics and Applications, Via Sommarive 14, 38123, Povo (TN), Italy\\
}

\email{\authormark{*}jacopo.terragni@unitn.it} 

\begin{abstract}
The material emitted from the target surface during laser ablation generates a net thrust (propulsion) in the opposite direction. The energetic efficiency of this laser-driven propulsion is given by the mechanical coupling coefficient ($C_\text{m}$). In this work we considered nanosecond UV laser ablation of the aluminum 6061 alloy to study the $C_\text{m}$ behaviour with different irradiating conditions. This is done by systematically changing: fluence, uniform/nonuniform intensity, and incident angle of the laser beam. In particular we found that, when dealing with nonuniform laser intensity, characterizing $C_\text{m}$ exclusively in terms of fluence is not fully satisfactory because the energy distribution over the irradiated area plays a key-role in the way material is removed -interplay between vaporization and phase-explosion- and thrust is generated. 
\end{abstract}

\section{Introduction}
Laser ablation (LA) is a well-established and widely used technique that finds application in various scientific and technological fields: from nano-material synthesis \cite{basso2019route} and surface modification \cite{lippert2005interaction} to chemical composition analysis \cite{pathak2012assessment}, from medicine \cite{gough2008laser} to inertial confinement nuclear fusion \cite{craxton2015direct}.

LA essentially consists in removing part of a solid material (target) by irradiation with intense laser energy. The ablation process can be briefly described as follows: laser optical energy is absorbed by the target which is heated to high temperature and consequently part of its material gets removed. For what concerns the mass removal process, three leading ablative mechanisms have been found: vaporization, phase-explosion, and melt flow \cite{fishburn2006study}. In the first two mechanisms material is actually removed from the target, while in the third it is only displaced from the irradiated area to the surrounding region without leaving the target. While vaporization occurs at every laser intensity, producing an expanding plume, phase-explosion requires an intensity threshold to be overcome. In fact, when laser intensity is high enough to bring the temperature close to its critical value $T \sim 0.9T_\text{c}$, the homogeneous nucleation and consequent growth of vapor bubbles inside the melt target results in a violent expulsion of liquid droplets, namely phase-explosion \cite{miotello1999laser, porneala2006observation}. 

Moreover, since the plume starts expanding in the early stages of LA, part of the laser intensity originally directed towards the target gets absorbed before reaching the surface, producing a partially ionized plume. This is the so-called plasma-shielding effect and it consistently affects the ablation efficiency. In summary, the outgoing material consists of an expanding plasma plume with eventually droplets of the molten material.

As an additional effect of LA, the outgoing ablated material generates a mechanical impulse on the target so that LA can be also used as a Propulsive technique (LAP)\cite{phipps2010laser}. Due to the previous considerations about mass removal mechanisms, only vaporization and phase-explosion actively take part in the thrust generation. In recent years, the opportunity to optically drive objects via LAP attracted great interest for space-based applications such as orbital debris mitigation \cite{phipps2014laser, lorbeer2018experimental, BATTISTON201736}, development of micro-thrusters \cite{phipps2004micropropulsion, karg2012laser}, and attitude control on nano-satellites \cite{jiao2014research}. 

When the propulsive aspects of LA are taken into account, one of the fundamental quantities that must be considered is the so-called mechanical coupling $C_\text{m}$, given by:

\begin{equation}
\label{eq:Cm}
    C_\text{m} = \frac{\int_0^\tau F_\text{t}\mathrm{d}t}{E}=\frac{J}{E}\qquad \left[\si{\newton\per\watt}\right],
\end{equation}
where $F_\text{t}$ is the generated thrust, $J$ is its impulse, $E$ is the laser energy, and $\tau$ is a later time with respect to the ablation process. Due to its definition, $C_\text{m}$ represents how efficient the mechanical impulse generation is in terms of the energy spent at this purpose. 

In this work we study how $J$ and $C_\text{m}$ change over different conditions of irradiation. In particular, we investigate their behaviour when either a uniform or nonuniform laser beam irradiates the target. 

\section{Experimental}

Laser ablation-generated thrust is here obtained by means of a \ce{KrF} pulsed excimer laser with a wavelength of \SI{248}{\nano\meter} and a pulse duration of \SI{20}{\nano\second}. 

\subsection{Laser beam intensity}

\begin{figure}[htbp]
\centering\includegraphics[width=\textwidth]{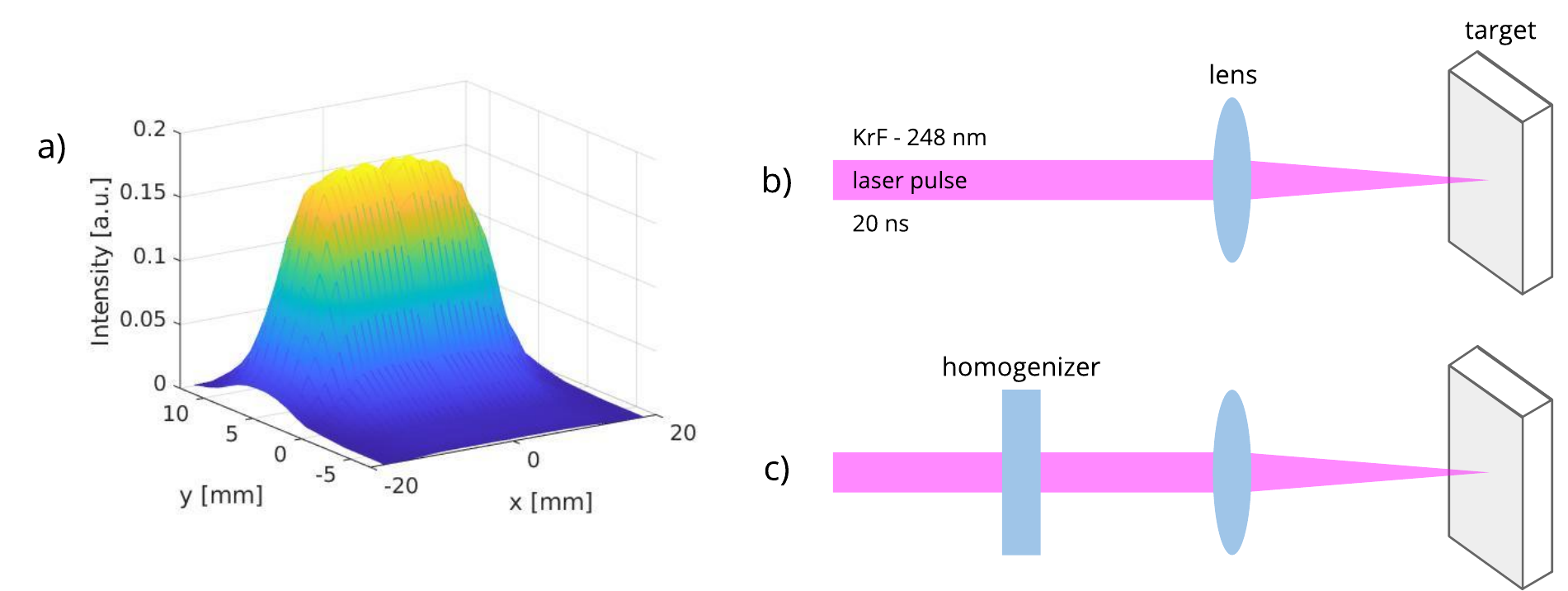}
\caption{\textbf{a)} Laser intensity profile. \textbf{b)} The laser beam is directly focalized onto the target. This optical path results in a nonuniform energy density distribution onto the irradiated area. \textbf{c)} The laser beam passes through a homogenizer before being focalized, resulting in a uniform energy density distribution onto the irradiated area.}
\label{fig:intensity_scheme}
\end{figure}

The spatial distribution of the laser pulse intensity is presented in Figure \ref{fig:intensity_scheme}a, where the $x$ and $y$ axes respectively represent the horizontal and vertical direction orthogonal to the laser propagation. It has been measured with the so-called knife-edge method \cite{deAraujo:09}. Knife-edge has been moved with \SI{1}{\milli\meter} steps both along the horizontal and the vertical direction. with a fixed laser energy of \SI{600}{\milli\joule} and a repetition rate of \SI{10}{\hertz}. For each knife-edge position, firing at \SI{10}{\hertz} with constant energy (\SI{600}{\milli\joule}), laser intensity was measured by means of a power meter: once a stable readout is reached, the average output intensity is acquired over an integration time of \SI{20}{\second}.

By looking at Figure \ref{fig:intensity_scheme}a, it can be seen that the intensity profile is nonuniform with nonsymmetric divergence (larger on the $x$-axis), as generally observed when dealing with excimer lasers \cite{endert1995excimer}. This laser pulse is then directed towards a solid target and then focalized on the target surface to increase its energy density in order to obtain ablation. 

In this work we considered two different configurations for its optical path: in the first one (Figure \ref{fig:intensity_scheme}b) the laser pulse is directly focalized onto the target by means of a spherical plano-convex lens, while in the second one (Figure \ref{fig:intensity_scheme}c) the laser pulse passes through a homogenizer (\texttt{Holo-Or RD 300} diffractive diffuser) before focalization. Different irradiation areas (spot), and consequently different optical energy density, have been studied simply by changing the relative distance between the lens and the target. All the considered spot areas have size of few \si{\milli\meter\squared} and they are reported in Section \ref{sec:results_discussion}.  
The difference between these two configurations in terms of the target ablation is discussed in Section \ref{sec:intensity_target}.

\subsection{Impulse measurements}

All the measurements presented in this work have been performed irradiating anticorodal aluminum (6061 series), an aluminum based alloy, which has been chosen due to its large usage in aerospace applications \cite{NGHIEP2018476,MOHITFAR2020155988}.

Impulse measurements are performed with a dedicated apparatus, which essentially consists in a ballistic pendulum enclosed in a vacuum chamber at pressures lower than \SI{E-4}{\milli\bar}. Impulses are generated by a single laser shot and measured by detecting the pendulum oscillations with a high speed camera. Technical details regarding this apparatus, along with its measurements procedure and its accuracy, are widely discussed in \cite{battocchio2020ballistic}.
The laser incidence angle, that is, the one between the beam propagation direction and the outgoing direction orthogonal to the target surface, is set to \ang{45}, except where otherwise stated.

To erase morphological irregularities or chemical contaminants that could affect the generated thrust, the surface of each target is polished first by means of a diamond paste and then with an ultrasonic bath. In addition to these procedures, once the target is mounted on the ballistic pendulum and set inside the vacuum chamber, its surface is laser-cleaned with a slightly focalized pulse ($F\le\SI{0.1}{\joule/\centi\meter\square}$). 

\section{Laser intensity onto the target}
\label{sec:intensity_target}

\begin{figure}[htbp]
\centering\includegraphics[width=\textwidth]{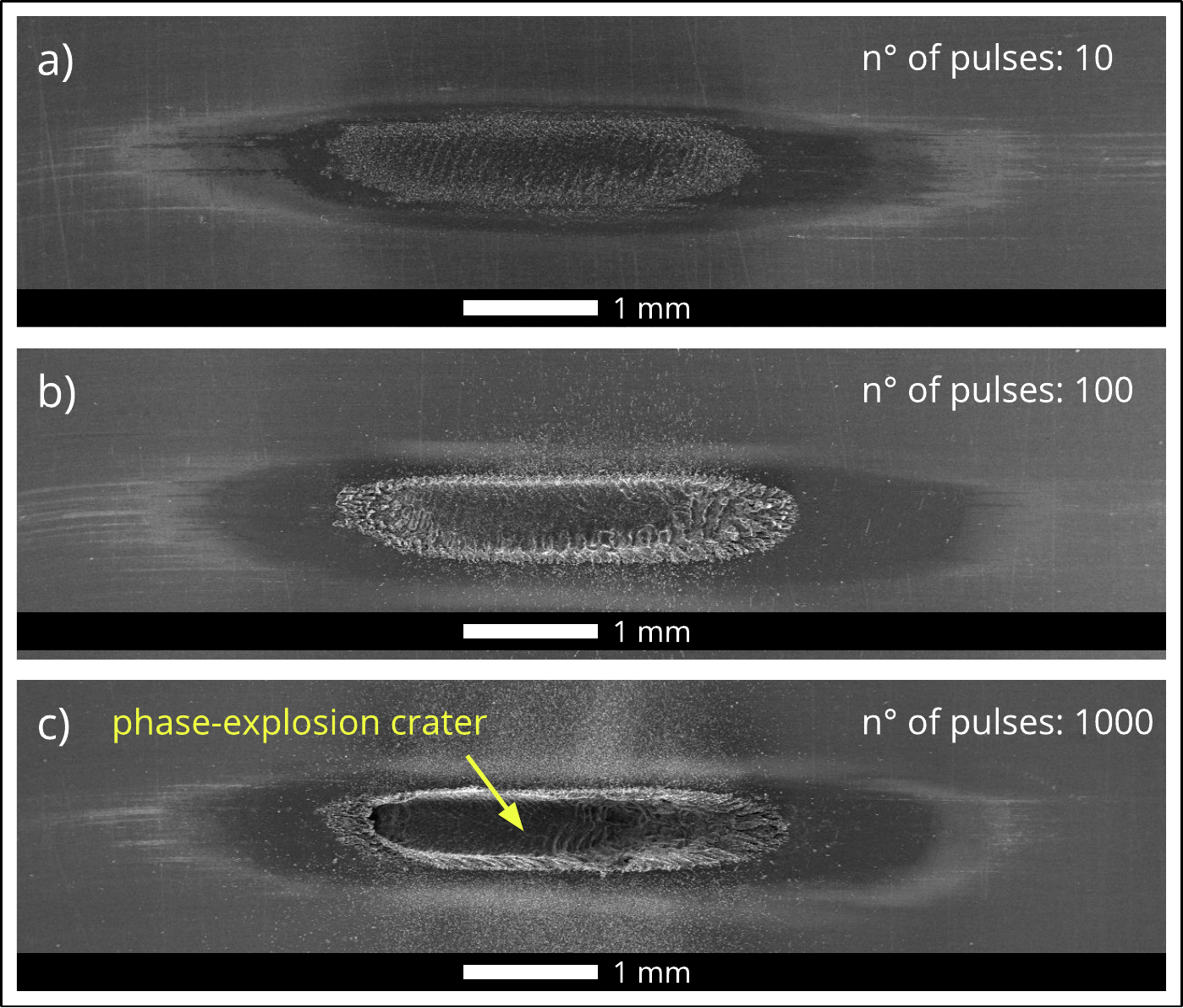}
\caption{Ablated spots after irradiation with \textbf{a)} 10, \textbf{b)} 100, and \textbf{c)} 1000 pulses of \SI{500}{\milli\joule} at \SI{10}{\hertz}. The optical path corresponds to the one depicted in Figure \ref{fig:intensity_scheme}b.}
\label{fig:no_diffuser_spot}
\end{figure}

When the configuration presented in figure \ref{fig:intensity_scheme}b is taken into account, the original laser pulse is directly focalized onto the target and the pulse intensity remains nonuniform over the irradiated surface.
Consequently, the optical energy per unit area over the irradiated spot (fluence) is definitely not constant and its mean value $F$, which is generally estimate as the ratio between energy and the spot area ($F=E/A$), may mislead in the interpretation of the ablation mechanism. In fact, local fluence in some regions of the irradiated spot may differ greatly from its average value, leading to a nonuniform heating of the target which may imply the concurrence of multiple regimes. As an example, we can consider the ablation shown in Figure \ref{fig:no_diffuser_spot}, where the three SEM images show the spot area after irradiation with 10 (a), 100 (b), and 1000 (c) pulses of \SI{500}{\milli\joule} at \SI{10}{\hertz}. 

The effects of different ablation regimes are clearly visible in Figure \ref{fig:no_diffuser_spot}, especially in (b) and (c) where the large number of pulses better highlights this aspect. The irradiated spot can be divided in two areas: one consisting in the large crater at the centre and one surrounding it. The former is due to phase-explosion and its position is consistent with the broaden peak position of the measured laser intensity profile (see Figure \ref{fig:intensity_scheme}a), while in the latter the only ablative mechanism is vaporization, since laser intensity is not enough to heat the target close to its critical temperature.

\begin{figure}[htbp]
\centering\includegraphics[width=0.6\textwidth]{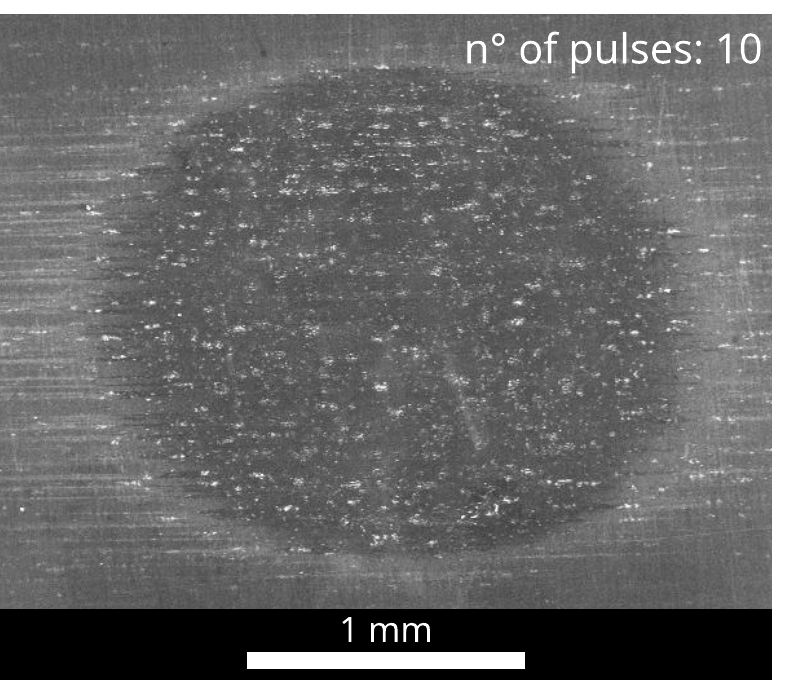}
\caption{Ablated spot after irradiation with 10 pulses of \SI{600}{\milli\joule} at \SI{10}{\hertz}. The optical path corresponds to the one depicted in Figure \ref{fig:intensity_scheme}c. The laser beam is here orthogonal to the target to better illustrate the role of the homogenizer.}
\label{fig:diffuser_spot}
\end{figure}

To avoid such inconvenience, a homogenizer can be inserted along the optical path, as depicted in Figure \ref{fig:intensity_scheme}c. This set-up results in a uniform energy density (top-hat profile of the laser intensity) on the ablated surface, so that the heating process of the target is the same within the spot, resulting in an homogeneous ablation as shown in Figure \ref{fig:diffuser_spot}. 

It is worth to notice that Figures \ref{fig:no_diffuser_spot}a and \ref{fig:diffuser_spot} represent the ablation effects induced by 10 pulses at approximately the same fluence ($F \simeq \SI{3.5}{\joule/\centi\meter\squared}$): in the first case phase-explosion occurs and a crater is produced, while in the second case there is no evidence of such ablative mechanism and material is exclusively removed by vaporization. The fact that the same fluence can lead to the different ablative regimes depending on the actual laser intensity profile is a key aspect of this work, because the ablation mechanism influences the thrust generation, which in turns determines the mechanical coupling as discussed in Section \ref{sec:results_discussion}. 



\section{Results and discussion}
\label{sec:results_discussion}

Before starting with the discussion of the impulse measurement results, an important preliminary consideration about the irradiated spot area can be stated: on one hand a larger spot could provide a higher impulse because it enlarges the target area from which material could be ablated, on the other hand a larger spot could result in lower impulse because a lower optical energy density would reduce the target heating and the consequent ablated mass flux (ablated mass per unit area per unit time). 

The size of the irradiated spot is an experimentally tunable parameter and then its twofold effect must be always take into account in any attempt to optimize the generated thrust. Among others, this aspect will be discussed in the following part of the paper, first by considering the nonuniform irradiation case (see Figure \ref{fig:intensity_scheme}b) and then the uniform one (see Figure \ref{fig:intensity_scheme}c).

\subsection{Impulse generated via nonuniform irradiation}
\label{sub:nonuniform}
\begin{figure}[ht!]
\centering\includegraphics[width=0.6\textwidth]{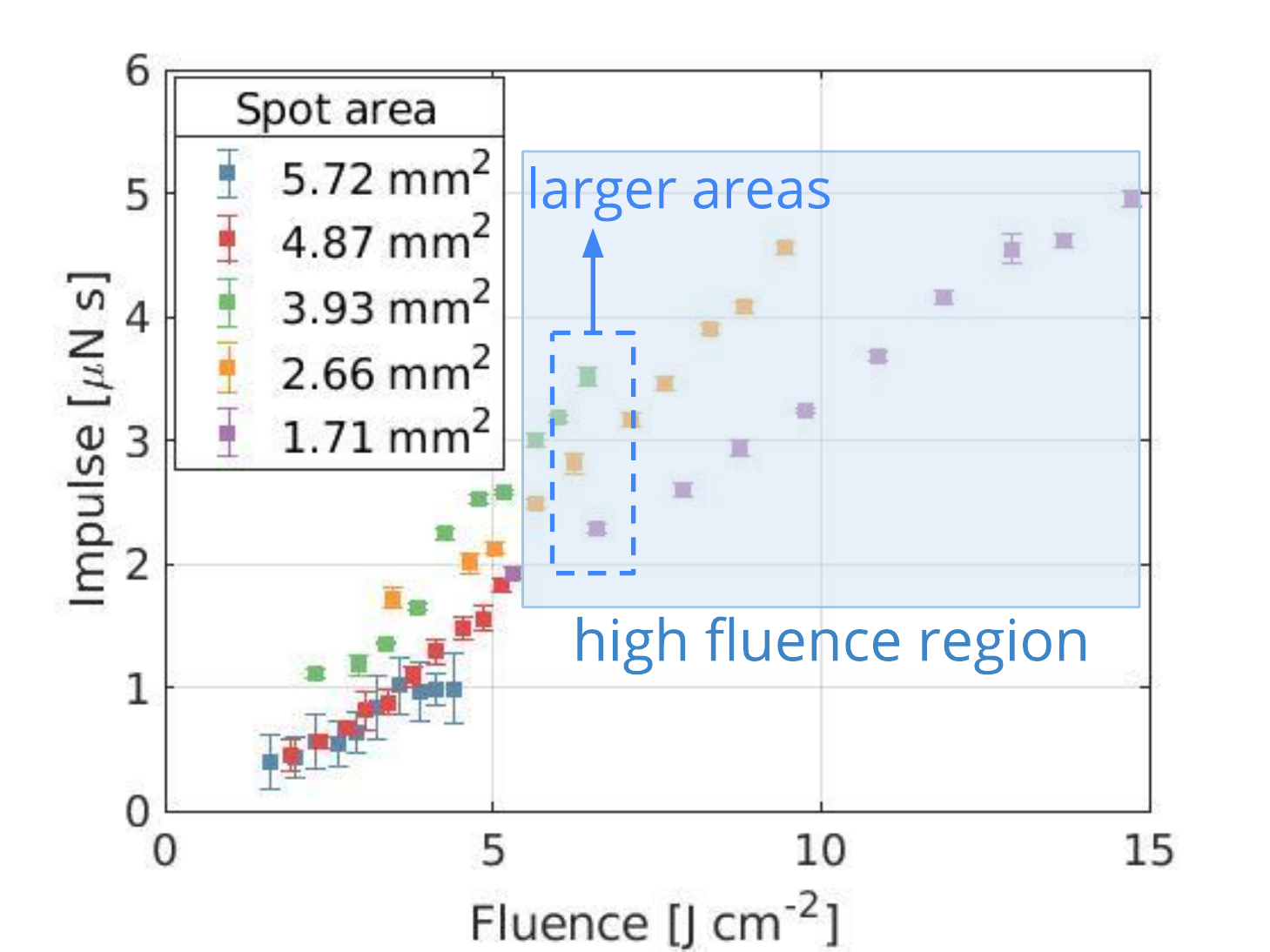}
\caption{Measured mechanical impulse as a function of fluence.}
\label{fig:impulse_fluence}
\end{figure}

The starting point of the discussion about impulse measurements results with nonuniform intensity is Figure \ref{fig:impulse_fluence}, in which the measured impulse is plotted as a function of fluence. 
First of all, it could be noticed that within a given spot area (color) the generated impulse exhibits a roughly linear behaviour with respect to fluence, similarly to what observed in \cite{zhou2020impulse}.

This figure can be divided into two regions: the high fluence region and the low fluence region, whose boundary can be set at around \SI{5}{\joule/\centi\meter\squared}. 
In the high fluence region (light blue-shaded area in Figure \ref{fig:impulse_fluence}), enlarging the spot area by defocusing the laser beam corresponds to increase the generated impulse, so that, for a given fluence, a higher impulse is generated by a larger spot area, as one can see by looking at the dashed-box region. Instead, a similar trend can't be extrapolated by looking at the low fluence region where the impulse-fluence relation appears to be quite tangled. 

\begin{figure}[ht!]
\centering\includegraphics[width=0.6\textwidth]{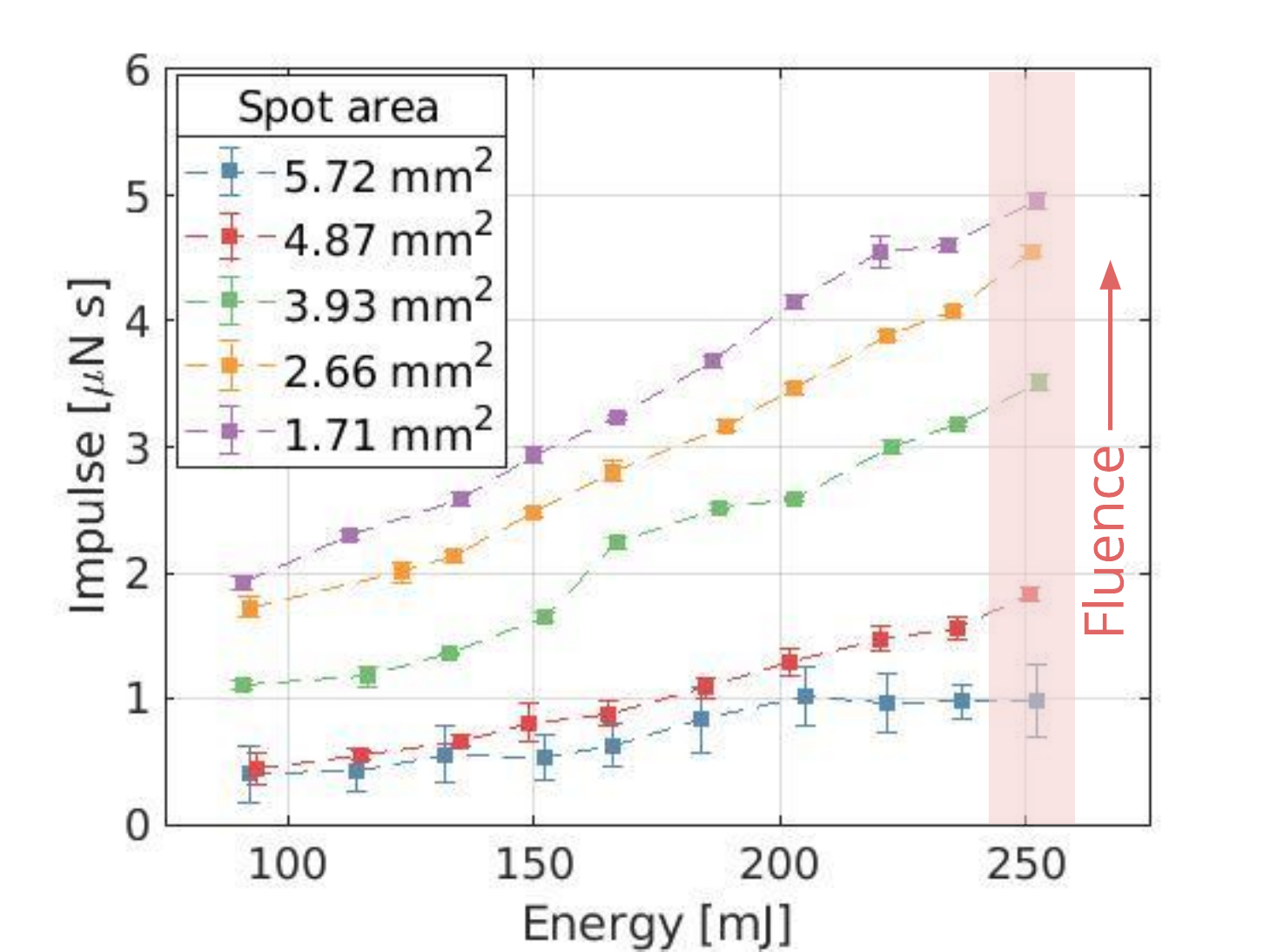}
\caption{Measured mechanical impulse as a function of the laser pulse energy.}
\label{fig:impulse_energy}
\end{figure}

To bring out some insights about this relation in the low fluence region, but also to extract additional considerations about the high fluence region, the measured impulse can be plotted against the laser pulse energy, as show in Figure \ref{fig:impulse_energy}.
It can be seen that, for a fixed energy (see for example the red-shaded area of Figure \ref{fig:impulse_energy}), a higher impulse is generated by a smaller spot area. This fact resolves which of the twofold aspects related to the spot size, discussed at the very beginning of this section, is the most relevant one. In fact, since a higher impulse is measured for a smaller spot area at fixed energy, the impulse growth due to increasing fluence is faster than the impulse decrease due to reducing the spot area. This is really important for what concerns the optimization of the thrust generation. Indeed, within the space of experimental parameters here considered, if the optical energy is fixed, the way to obtain the best thrust performance is to focalize onto the smallest area. 

\begin{figure}[ht!]
\centering\includegraphics[width=0.6\textwidth]{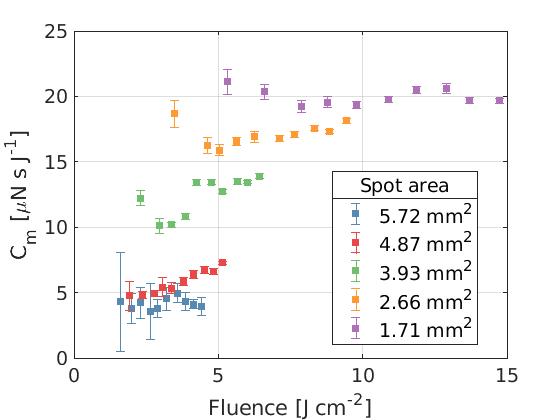}
\caption{Mechanical coupling ($C_\text{m}$) as a function of fluence.}
\label{fig:cm_fluence}
\end{figure}

The next step in our discussion is to consider how the mechanical coupling $C_\text{m}$ behaves with respect to fluence, which is shown in Figure \ref{fig:cm_fluence}, from which various useful considerations can be extracted.

First of all, it is worth to notice that $C_\text{m}$ measured values are in good agreement (within the same order of magnitude) with other results reported in literature, although obtained with some different experimental conditions (laser wavelength \cite{yu2020experimental,Tsuruta2014}, pulse duration \cite{zhou2020impulse,Tsuno20}, pressure \cite{tran2017impulse,senegavcnik2020propulsion}, \dots).
Another extremely important aspect is that, remaining within the space of our experimental parameters, $C_\text{m}$ is not uniquely characterized by the irradiating fluence. This means that different $C_\text{m}$ values can be obtained at a given fluence. 

\begin{figure}[ht!]
\centering\includegraphics[width=0.6\textwidth]{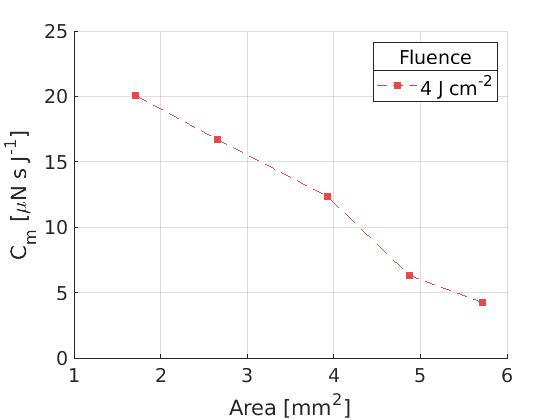}
\caption{Mechanical coupling ($C_\text{m}$) as a function of the spot area at the fixed fluence of $F=\SI{4}{\joule/\centi\meter\squared}$. $C_\text{m}$ values are extracted from linear interpolation of the measured values shown in Figure \ref{fig:cm_fluence}.}
\label{fig:cm_area}
\end{figure}

To stress out this point, Figure \ref{fig:cm_area} depicts the case when $F=\SI{4}{\joule/\centi\meter\squared}$. 
It shows how $C_\text{m}$ significantly reduces (up to nearly 25\% of its value) moving from the most to the least focalized laser beam. Then, when ablation is performed with nonuniform laser intensity, the energetic efficiency of the generated mechanical impulse does not simply depends on the ratio $E/A$, but also on how this ratio is obtained. 

\begin{figure}[htbp]
\centering\includegraphics[width=0.6\textwidth]{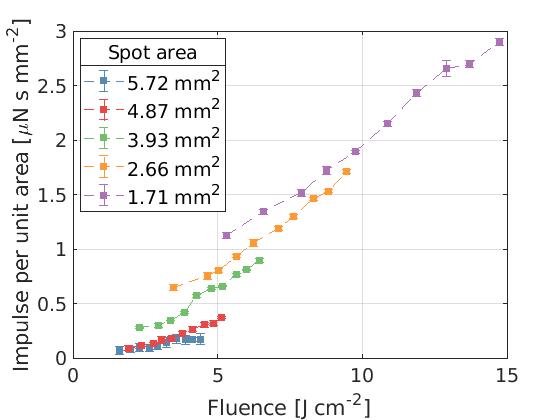}
\caption{Mechanical impulse per unit area as a function of fluence.}
\label{fig:impulsepua_fluence}
\end{figure}

An important step forward in the analysis of these results would be to connect the measured impulse and its efficiency to the ablation mechanisms (vaporization and phase-explosion). This aspect is not straightforward in nonuniform laser intensity conditions, because the energy density over the irradiated area is not constant and the ablation of a subarea of the spot proceeds according to the energy density impinging on that given subarea. Few insights can be obtained by looking at Figure \ref{fig:impulsepua_fluence}, where the generated impulse per unit area, obtained by considering the same spot area used to calculate laser fluence, is plotted against fluence. 

At a given fluence, higher values are obtained with smaller areas. Some steps in impulse per unit area, with the exception of the two largest spots, can be seen also in Figure \ref{fig:impulsepua_fluence} and their presence could be also related to the nonuniform intensity of the laser pulse that allows both vaporization and phase-explosion to simultaneously take part in ablation. In fact, for the two largest areas, where only vaporization is present because even the energy density in the center of the spot is not enough to reach the phase-explosion onset, the two curves (blue and red) merge together. On the contrary, when both ablation mechanisms are present, the impulse over irradiated area ratio results differently for different areas at the same fluence because their contribution in generating impulse is different. In particular, larger impulses per unit area at fixed fluence are obtained  with smaller spots. This may suggest that focusing the laser beam increases the ratio between the area interested by phase-explosion and the overall irradiated area \cite{TSURUTA201746}.

\begin{figure}[ht!]
\centering\includegraphics[width=0.6\textwidth]{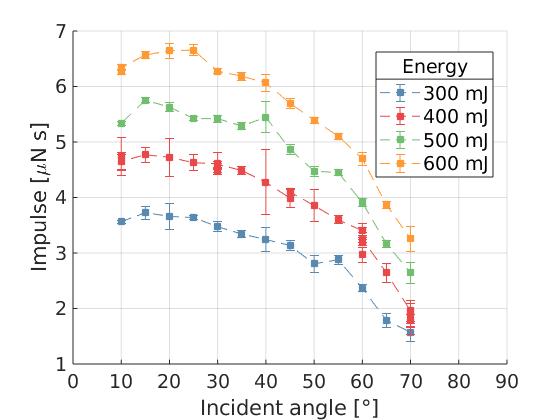}
\caption{Mechanical impulse as a function of the incident angle.}
\label{fig:impulse_angle}
\end{figure}

Moving the lens position with respect to the target is not the only way to change the size of the irradiated area and investigate its role. It can be also varied by tilting the target with respect to the laser beam, so that the irradiated area for an incident angle $\theta$ is given by
\begin{equation}
\label{eq:inc_angle}
    A(\theta) = \frac{A_0}{\cos(\theta)},
\end{equation}
where  $A_0 = A(0)$ is the spot area measured at normal incidence.
Figure \ref{fig:impulse_angle} shows impulses obtained with respect to $\theta$ for different pulse energies. Since each angle defines a different spot area according to \eqref{eq:inc_angle}, curves obtained for larger energies are shifted upwards because increasing energy corresponds to increasing fluence, which implies a higher impulse.

All four curves depicted in Figure \ref{fig:impulse_angle} show the same two features: the generated impulse tends to zero for large angles while a maximum is observed for small angles. The former is due to the fact that a larger angle corresponds to a larger area and, since energy is fixed for each curve, to a smaller fluence, which in turn implies a smaller generated impulse. The latter is caused by the plasma-shielding effect \cite{zhao2016influence,wang2017laser} which depends, among other factors, directly on the beam optical path length inside the plume. In fact, since this optical path increases for small incident angles, the shielding becomes more effective and reduces the laser fluence on the target.

\begin{figure}[ht!]
\centering\includegraphics[width=0.6\textwidth]{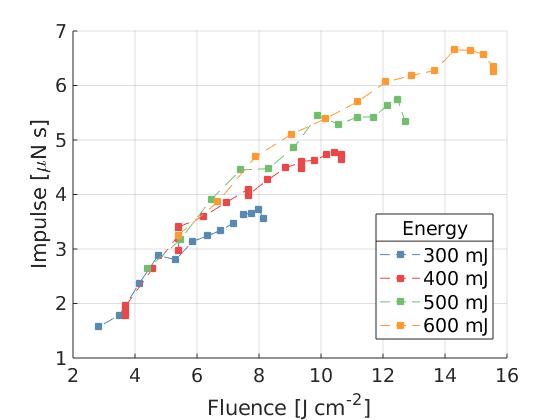}
\caption{Mechanical impulse as a function of fluence due to the variation of the incident angle.}
\label{fig:impulse_anglefluence}
\end{figure}

In Figure \ref{fig:impulse_anglefluence} the generated impulse is shown as a function of fluence, where areas have been computed according to \eqref{eq:inc_angle}. Since the curves have been obtained at a given energy, $C_\text{m}$ will share the same behaviour. Such curves look different from what discussed in Figure \ref{fig:cm_fluence}, suggesting once again that the $C_\text{m}$ behaviour with respect to fluence depends on how the energy-area ratio is produced.

\begin{figure}[htbp]
\centering\includegraphics[width=0.6\textwidth]{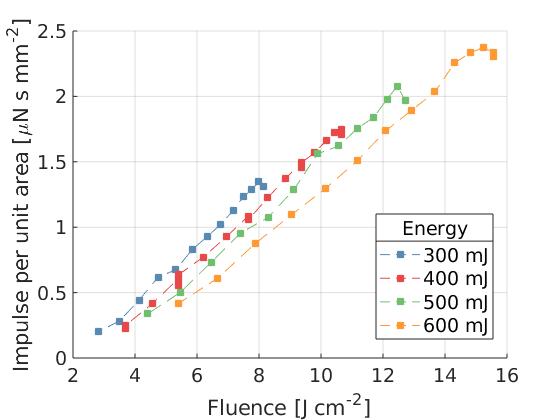}
\caption{Impulse per unit area as a function of fluence due to the variation of the incident angle.}
\label{fig:impulsepua_anglefluence}
\end{figure}

Measured impulse could eventually be normalized by the spot area, as shown in Figure \ref{fig:impulsepua_anglefluence}. Once again steps among curves can be detected, and curves at lower energy show a higher impulse per unit area at given fluence, recovering the same behaviour on Figure \ref{fig:impulsepua_fluence}.

As a conclusive remark about this discussion about the impulse generation in nonuniform intensity conditions, we can affirm that the energetic efficiency of the generated impulse is clearly a function of the optical mean energy density directed towards the target, but this is not enough: a $C_\text{m}$ value in only properly characterized if both the actual laser intensity profile and how the energy-area ratio is obtained are specified.

\subsection{Impulse generated via uniform irradiation}
\begin{figure}[htbp]
\centering\includegraphics[width=0.6\textwidth]{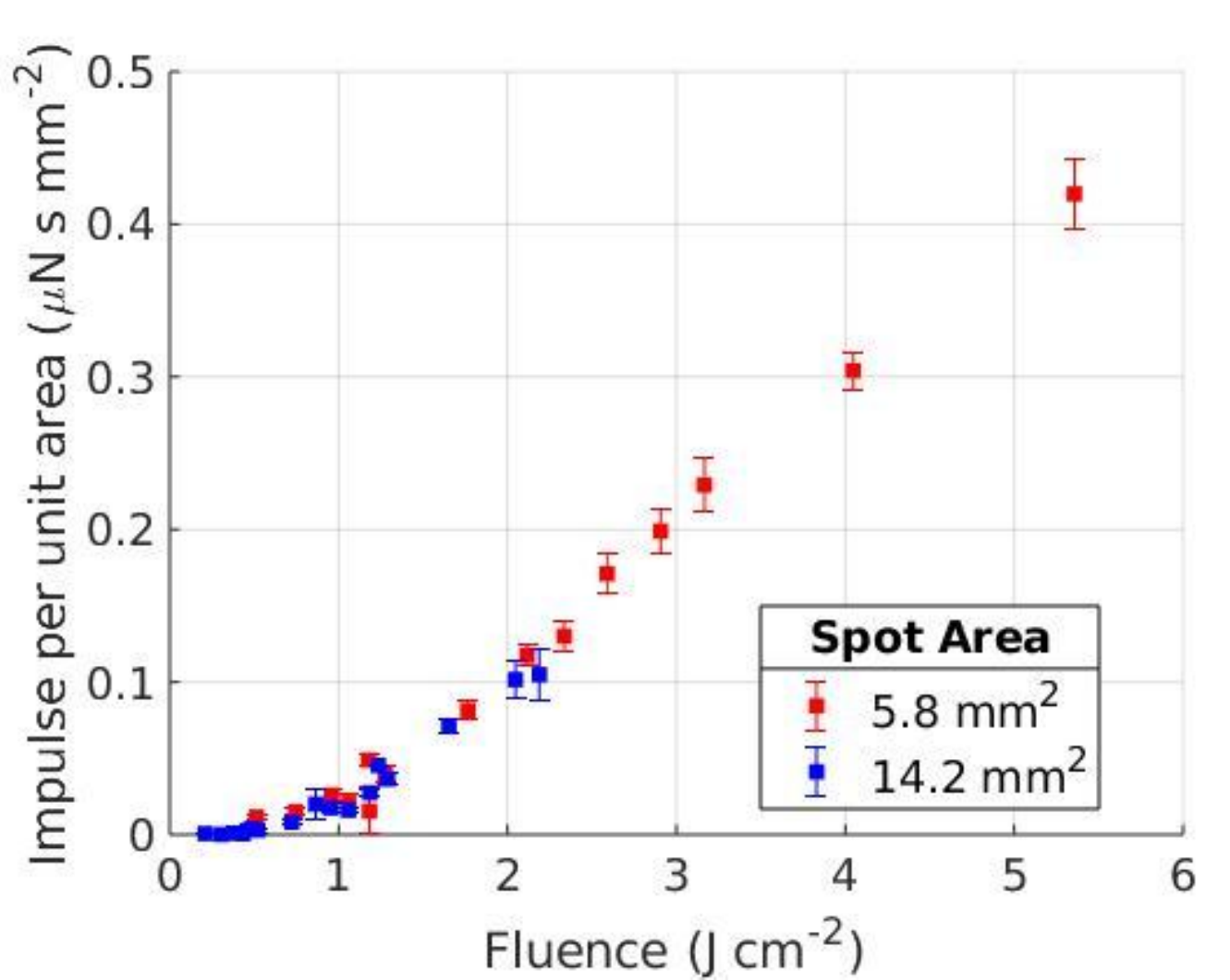}
\caption{Impulse per unit area as a function of fluence with uniform conditions of irradiation.}
\label{fig:impulsepuadiff_fluence}
\end{figure}

It is evident from what discussed in Section \ref{sub:nonuniform} that the only way to properly relate the impulse generation to the ablation mechanisms involved is performing ablation with an optical energy density uniformly distributed over the target. To fulfill this purpose, a homogenizer have been inserted in the optical path, as depicted in Figure \ref{fig:intensity_scheme}c. 
Our results with this experimental setup are presented in Figure \ref{fig:impulsepuadiff_fluence}. 
Here the impulse per unit area is plotted against fluence. The two sets of data exhibit a single trend over the all fluence domain. Having a common trend in the generated impulse per unit area implies that their difference in the generated impulse is exclusively determined by their different area. This in turns implies that, at a given fluence, the heating process of the irradiated area is the same in both cases, without any further dependence on the spot size. 

Moreover, in the fluence region when there is superposition of two sets of data ($F<\SI{3}{\joule/\centi\meter\squared}$), the ablation mechanism involved is vaporization. This comes by looking at Figure \ref{fig:diffuser_spot}, where no evidence of phase-explosion are present at $F\simeq\SI{3.5}{\joule/\centi\meter\squared}$. In addition to that, it can be noticed that, by comparison of Figure \ref{fig:impulsepuadiff_fluence} with Figure \ref{fig:impulsepua_fluence} (blue and red curves), the impulse per unit area measured values are in good agreement in the two cases for $F<\SI{5}{\joule/\centi\meter\squared}$. These last two facts suggest a final additional consideration: the dominating ablation mechanism involved in the blue and red curve (largest areas of Figure \ref{fig:impulsepua_fluence}) is actually vaporization, as stated in the previous section about Figure \ref{fig:impulsepua_fluence}.  

\section{Conclusions}

In this work we performed a study of the ns-UV laser ablation-generated mechanical impulse and its coupling coefficient ($C_\text{m}$) by systematically changing the experimental conditions of target irradiation. The first conclusion we can draw is that, when a nonuniform laser beam is used to perform ablation, the energetic efficiency of generated impulse is not uniquely determined by the irradiating fluence. Within the presented experimental conditions, we found that the best $C_\text{m}$ result at fixed fluence is obtained by reducing the irradiated spot area. We also found that $C_\text{m}$ behavior is different if fluence is varied by focalizing/defocalizing the laser beam or changing the incident angle between the beam and the target. Finally, we gained insights about the importance of a uniform energy density irradiation when one aims to connect the energetic efficiency of the impulse generation with the ablation mechanisms involved.

\begin{backmatter}
\bmsection{Funding}
This research was partially supported by the Istituto Nazionale di Fisica Nucleare (INFN), within the project Glare-X.


\bmsection{Disclosures} The authors declare no conflicts of interest.

\end{backmatter}

\bibliography{AppOptBibliography}






\end{document}